\documentclass[lettersize,journal]{IEEEtran}
\usepackage{amsmath,amsfonts}
\usepackage{algorithmic}
\usepackage{algorithm}
\usepackage{array}
\usepackage[font=footnotesize, labelfont=bf, justification=raggedright, singlelinecheck=false]{caption}
\usepackage{subcaption}
\usepackage{textcomp}
\usepackage{stfloats}
\usepackage{url}
\usepackage{multirow}
\usepackage{makecell}
\usepackage{diagbox}
\usepackage{verbatim}
\usepackage{graphicx}
\usepackage{cite}

\begin{document}

\title{Toward Multi-Functional LAWNs with ISAC: Opportunities, Challenges, and the Road Ahead}


  \author{
    \IEEEauthorblockN{Jun Wu,  Weijie Yuan, \IEEEmembership{Senior Member, IEEE}, Xiaoqi Zhang,  Yaohuan Yu, 
    Yuanhao Cui, \IEEEmembership{Member, IEEE},\\
    Fan Liu, \IEEEmembership{Senior Member, IEEE},
    Geng Sun, \IEEEmembership{Senior Member, IEEE},
    Jiacheng Wang, \IEEEmembership{Member, IEEE},\\
    Dusit Niyato, \IEEEmembership{Fellow, IEEE}, and 
    Dong In Kim, \IEEEmembership{Life Fellow, IEEE}
  }
  \thanks{

  J. Wu, W. Yuan, and Y. Yu are with the School of Automation and Intelligent Manufacturing, Southern University of Science and Technology, Shenzhen 518055, China (email: wuj2021@mail.sustech.edu.cn; yuanwj@sustech.edu.cn; yuyh2024@mail.sustech.edu.cn).
  X. Zhang is with the School of Electrical and Data Engineering, University of Technology Sydney, Sydney, 2007, Australia (e-mail: Xiaoqi.Zhang@student.uts.edu.au)
  Y. Cui is with the School of Information and Communication Engineering, Beijing University of Posts and Telecommunications, Beijing 100876, China (e-mail: cuiyuanhao@bupt.edu.cn).
  F. Liu is with the National Mobile Communications Research Laboratory, Southeast University, Nanjing 210096, China  (e-mail: fan.liu@seu.edu.cn).
  G. Sun is with the College of Computer Science and Technology, Jilin University, Changchun 130012, China (E-mail: sungeng@jlu.edu.cn).
J. Wang and D. Niyato are with the College of Computing and Data Science, Nanyang Technological University, Singapore 639798 (e-mails:  jiacheng.wang@ntu.edu.sg; dniyato@ntu.edu.sg).
D. I. Kim is with the Department of Electrical and Computer Engineering, Sungkyunkwan University, Suwon 16419, South Korea (e-mail: dongin@skku.edu).

  }
  }
\maketitle

\begin{abstract}
Integrated sensing and communication (ISAC) has been envisioned as a foundational technology for future low-altitude wireless networks (LAWNs), enabling real-time environmental perception and data exchange across aerial-ground systems. In this article, we first explore the roles of ISAC in LAWNs from both node-level and network-level perspectives. We highlight the performance gains achieved through hierarchical integration and cooperation, wherein key design trade-offs are demonstrated. Apart from physical-layer enhancements, emerging LAWN applications demand broader functionalities.  To this end, we propose a multi-functional LAWN framework that extends ISAC with capabilities in control, computation, wireless power transfer, and large language model (LLM)-based intelligence. We further provide a representative case study to present the benefits of ISAC-enabled LAWNs and the promising research directions are finally outlined. 
\end{abstract}


\section{Introduction}

\begin{figure}[t]
    \centering
    \includegraphics[width=\linewidth]{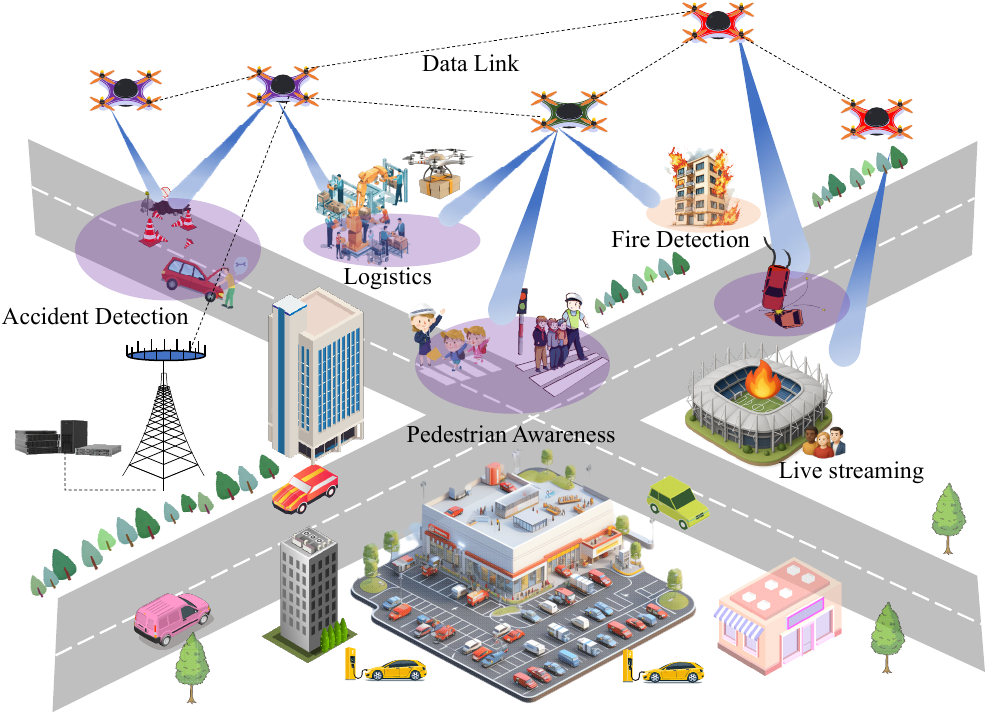}
    \caption{ISAC-enabled LAWN applications, e.g., accident detection, fire detection, logistics, pedestrian awareness, and live streaming in urban environments, providing flexible and efficient service coverage.}
    \label{figsceISAC} \vspace{-0.5cm}
\end{figure}

The rapid development of low-altitude economic activities, ranging from urban drone logistics to intelligent emergency response, has posed significant challenges to traditional wireless infrastructure designs. To meet the complex demands of these emerging applications, low-altitude wireless networks (LAWNs) must evolve to support dense aerial deployments with integrated capabilities for reliable low-latency communication and accurate environmental sensing \cite{11017717}.
In this context, integrated sensing and communication (ISAC) emerges as a key enabler, offering a unified physical-layer framework that simultaneously supports environmental perception and data exchange. By breaking the traditional separation between sensing and communication (S\&C), ISAC enhances spectral efficiency, reduces hardware redundancy, and enables context-aware interactions between distributed aerial-ground LAwNs \cite{9737357}. As a result, ISAC becomes a foundational technology for realizing scalable and intelligent LAWN operations, as depicted in Fig. \ref{figsceISAC}.

However, as LAWN operations evolve toward increasingly sophisticated and autonomous paradigms, the foundational advantages of ISAC, while critical, are no longer sufficient to accommodate the multifaceted demands of real-world aerial missions. Emerging LAWN applications will transcend the conventional S\&C functionalities, necessitating efficient incorporation of dynamic real-time control, distributed edge intelligence, and wireless power transfer (WPT) \cite{yuan2025_ground_to_sky}. Specifically, \textit{control-aware} integration leverages ISAC-derived real-time sensing and communication to stabilize formation flight, ensure precise trajectory tracking, and maintain robust attitude control. In parallel, \textit{computation-aware} integration exploits ISAC-informed channel and environment knowledge to optimize task allocation and computation offloading at the edge, with large language models (LLMs) further enriching edge intelligence through semantic reasoning and adaptive decision-making. Meanwhile, \textit{energy-aware} integration capitalizes on adaptive and directional WPT to sustain ISAC transceivers and prolong mission endurance. The interplay between ISAC and these functionalities is pivotal to enabling fully autonomous drone operations in terms of extending mission endurance and sustaining reliable control in resource-constrained aerial environments. To this end, we introduce a forward-looking architectural and operational visions that define wireless infrastructures not only as  S\&C enablers, but also as unified platforms for control, computation, large language model (LLM)-based decision-making, and energy delivery. Motivated by this transformative architectural shift, this article conducts a systematic examination of the opportunities and open challenges in advancing LAWNs from traditional ISAC systems to a multi-functional paradigm. The main contributions are summarized as follows 
\begin{itemize}
    \item We begin by presenting core benefits and architectures of ISAC-based LAWNs, emphasizing the node/network-level ISAC design challenges and tradeoffs. 
    \item Subsequently, we explore critical functional extensions beyond ISAC, including wireless control, distributed computation offloading, powering, and LLM-based intelligence, detailing the emerging methodologies that enable these capabilities to be efficiently integrated into the LAWN infrastructure. 
    \item Finally, we outline promising directions, i.e., aircraft traffic management, identity authentication, and privacy preservation accelerating the development of LAWNs.
\end{itemize}

\section{ISAC-Enabled LAWN}

In this section, we explore how ISAC enhances the performance of LAWNs at both the node and network levels. At the node level, ISAC enables individual aerial devices to leverage sensing outputs to refine communication functions, and vice versa. At the network level, it facilitates global coordination among distributed nodes, supporting collaborative perception and control under dynamic topologies.

\subsection{Node-Level ISAC}
The node-level ISAC empowers individual LAWN participants, e.g., various drones and ground users, to simultaneously exploit S\&C functions within a unified transceiver architecture, and thereby, each node is capable of autonomously adapting to dynamic environments. We elaborate on the core challenges corresponding to the node-level ISAC design.
\subsubsection{Joint Waveform Design}

A fundamental advantage of ISAC lies in the integration gain enabled by unified waveform design. Instead of treating S\&C as separate subsystems, a shared waveform allows for hardware reuse and reduced spectral overhead. However, unlike terrestrial ISAC scenarios that typically assume quasi-static environments, LAWN systems must contend with rapidly time-varying channels and complex multipath propagation in unstructured air-ground settings. In this context, conventional orthogonal frequency division multiplexing (OFDM) waveforms often suffer severe performance degradation under significant Doppler shifts. To address these challenges, several candidate waveforms have been proposed for dual-functional use in LAWNs. Orthogonal time frequency space (OTFS) modulation \cite{9508932} provides a delay-Doppler domain representation that is resilient to high-mobility effects, making it well-suited for low-altitude aerial platforms.
Likewise, affine frequency division multiplexing (AFDM)\cite{10087310}, by mapping signals over an affine-transformed grid, enhances robustness against doubly dispersive channels and offers design flexibility tailored to non-stationary environments. In addition, hybrid schemes that embed sensing pilots into communication frames or superimpose auxiliary sensing signals onto data streams enable waveform-level fusion while minimizing mutual interference.  More recent efforts focus on task-driven waveform optimization, such as minimizing sensing mean squared error under communication quality-of-service (QoS) constraints, or vice versa. 
By enabling S\&C to coexist harmoniously within a shared waveform, such designs are capable of offering foundational support for LAWN nodes.

\subsubsection{Sensing-assisted Communication}

One of the primary obstacles to reliable transmission within LAWNs is the timely acquisition of accurate channel state information (CSI). Conventional solutions, which rely heavily on pilot transmission and feedback protocols, incur non-negligible signaling overhead and consume portions of spectrum and power resources. However, the pilot-based scheme may suffer from performance degradation under high-mobility conditions with dynamically changing environments, where the coherence time of the channel is severely reduced, rendering pilot-based CSI quickly outdated. Moreover, the limited computational capacity and energy budget of drones make frequent feedback and beam training infeasible. Sensing-assisted communication (SaC) is capable of offering a promising alternative. By leveraging the radar sensing waveforms for environmental perception, it becomes possible to infer essential channel parameters, e.g.,  position, velocity, and angle-of-arrival (AoA) of the communication counterpart, without relying on dedicated pilots or feedback mechanisms. Subsequently, the extracted location and motion state of the receiver allow the transmitter to perform precise beam tracking and prediction by incorporating previous measurements, ensuring spatial alignment without exhaustive beam search procedures. 

Beyond beam management, SaC also plays a pivotal role in dynamic resource allocation and physical-layer security enhancement. On the one hand, by capturing channel variations and user mobility, SaC facilitates proactive and context-aware resource allocation. For example, idle nodes can leverage sensing signals to detect and estimate user states, enabling low-latency handover without the need for explicit signaling exchange. Furthermore, SaC allows bandwidth and beamwidth resources to be dynamically adjusted in accordance with both communication QoS requirements and sensing resolution constraints, especially under spectrum reuse and multi-user interference conditions. In terms of security enhancement, SaC can be aware spatial information about potential eavesdroppers, such that the transmitter can employ directional transmission or artificial noise to suppress information leakage while maintaining sensing accuracy \cite{10168298}. 

\subsubsection{Communication-assisted Sensing}

To further unlock the potential of LAWN, cmmunication-assisted sensing (CaS) is capable of addressing the core challenge of real-time environmental perception. Conventional methods typically rely on dedicated sensing hardware and tightly synchronized transceiver pairs. However, the addition of sensing modules increases energy consumption, while maintaining synchronization across multiple aerial nodes remains difficult in dynamic airspace environments. In contrast, CaS reuses communication waveforms for sensing, such that each LAWN node extracts essential parameters directly from the communication channel including range and AoA, thereby eliminating the need for dedicated radar waveforms and substantially reducing signal processing complexity. 

In addition, CaS enables beyond individual field-of-view awareness through the uplink transmission of sensing results to remote receivers. In this mechanism, the sensing node gathers local observations and forwards them to other terminals that are unable to sense the target directly. To further improve sensing resolution, synthetic aperture radar (SAR) imaging can be functionally embedded into the CaS framework. Unlike instantaneous reflection-based estimations, SAR provides substantial gain by coherently accumulating reflected signals over synthetic trajectories, effectively emulating a large physical aperture for high-resolution sensing. By exploiting the spatial diversity introduced by predetermined drone trajectories, synthetic aperture can be formed through uplink communication signals, effectively circumventing the need for additional radar-specific subsystems.
To name a few, the low-altitude security monitoring can be achieved by CaS-based schemes, where a base station (BS) may detect unauthorized drones via SAR imaging and subsequently transmit the processed results to authorized drones navigating nearby. This allows the authorized drones to respond to potential threats that are outside their own sensing field of view, despite lacking a direct line-of-sight to authorized drones, e.g., the intruders. 
However, a fundamental challenge in devising such a CaS system lies in the tradeoff between communication rate and sensing distortion. Unlike remote estimation, where raw observations are explicitly encoded for transmission, CaS primarily operates by conveying processed sensing results whose distortion is jointly contingent upon the local inference quality and the foundational waveform structure. Designs prioritizing spectral efficiency for communication may uninevitably compromise sensing's perceptual resolution, and conversely, sensing-centric configurations can degrade data throughput. These conflicting objectives give rise to a nontrivial, task-dependent rate–distortion region, whose characterization under high-mobility and multi-agent settings remains a pivotal open challenge for future LAWN system design \cite{10845869}.

\subsection{Network-Level ISAC}
While node-level ISAC enables each LAWN node to locally exploit S\&C synergy, its benefits remain constrained without coordination at the network scale. In large-scale LAWN deployments, where multiple heterogeneous nodes operate concurrently under shared spectrum and dynamic topology, it becomes essential to extend ISAC design to the network level, which addresses the collective challenges of interference coupling, resource competition, and information consistency across nodes, thereby supporting scalable and collaborative system behavior.
\subsubsection{Interference Management}

The coexistence of heterogeneous LAWN nodes operating over shared spectral and spatial domains introduces a complex interference environment that challenges conventional system design paradigms. Unlike traditional wireless networks, where interference is primarily treated as a communication impairment, ISAC systems demand a dual-domain perspective, in which the same interference simultaneously impacts both communication throughput and sensing accuracy, manifesting in false alarm rates and diminished range-Doppler resolution. This coupling becomes particularly severe in dense deployments and high-mobility LAWNs \cite{meng2025cooperative}, where beam pattern overlap and waveform interaction in both spatial and temporal dimensions exacerbate mutual interference. To effectively mitigate these challenges, interference management must evolve from isolated link-level techniques to coordinated optimization frameworks. A possible direction involves geometry-aware drone deployment and trajectory planning, wherein mobile drones dynamically adjust their flight path to reshape the interference topology. By fully leveraging spatial diversity and angular separation, it is capable of reducing line-of-sight conflicts with the served ground units or targets, and thereby suppressing cross-node interference in both S\&C.  In general, the spatiotemporal coordination can be further enhanced by jointly optimizing beamforming vectors and dual-functional waveforms, enabling drones to steer their main lobes toward intended directions while projecting their sidelobes into the orthogonal complement of the interference subspace. Apart from mitigating interference, network-level ISAC systems can harness constructive interference (CI) through collaborative sensing and data fusion, effectively transforming traditionally undesirable signal components into valuable information. To gain these benefits, however, the system must maintain precise inter-node coordination. Toward this end, we proceed to examine centralized and distributed ISAC design strategies, which facilitate scalable coordination across heterogeneous LAWN nodes.

\subsubsection{Centralized/Distributed Networked-ISAC Design}

In large-scale LAWN deployments, centralized ISAC design offers a structured framework wherein heterogeneous nodes transmit their sensing observations to a designated fusion center via dedicated feedforward or feedback links. This architecture enables global situational awareness and facilitates system-wide decision-making, e.g., LAWN-enabled collaborative positioning, where the fusion center exploits fully aggregated observations to achieve enhanced localization performance. However, the reliance on a centralized mechanism introduces critical vulnerabilities. Failures, congestion, or security breaches at the fusion node can propagate throughout the entire network, potentially leading to widespread degradation or even complete system collapse, thereby severely impairing both sensing accuracy and communication reliability. To circumvent this, a transition toward distributed ISAC designs tailored for the LAWN emerges as a resilient alternative. Instead of relying on centralized processing, distributed architectures allow each node to maintain local inference capabilities and collaborate with neighbors through limited drone-to-drone (D2D) communication and thus, enhancing scalability and robustness in dynamic environments where continuous connectivity to a fusion center cannot be guaranteed.

 Distributed ISAC relies on consensus-based mechanisms to align local estimates across the network by leveraging message passing and decentralized inference approaches. Through iterative information exchange, LAWN agents gradually converge to globally consistent decisions while preserving flexibility under changing topologies. However, the environmental uncertainty and sensing imperfections prevalent in LAWNs, such as multi-path reflections and clutter interference, often lead to non-Gaussian observation distributions that pose a challenge to traditional consensus algorithms based on linear or unimodal models. To address this, some advanced methods have been well-investigated, e.g., Gaussian mixture-based distributed (GMD) estimation frameworks \cite{10614394}, which model local beliefs as Gaussian mixtures, capturing complex uncertainty structures in a compact probabilistic form. Through message passing of mixture parameters rather than original measurements, LAWN agents can iteratively update their posterior distributions, achieving robust inference with limited bandwidth usage and low latency. For instance, consider a distributed LAWN-enabled multi-target tracking scenario, where each LAWN agent detects partial and noisy observations of moving targets. GMD-based belief exchange enables the network to jointly reconstruct the dynamic target states across space and time, even in the presence of occlusions or intermittent connections. 
 \subsection{Lessons Learned} Building upon the aforementioned discussions, the synergy of node-level and network-level ISAC  not only improves spectral utilization but also fully exploits spatial diversity to satisfy various S\&C requirements. The main comparisons between node-level and network-level ISAC designs are summarized in TABLE \ref{table:ISAC_comparison}. Nevertheless, despite several benefits of ISAC for LAWN systems, the increasing complexity of real-world applications demands capabilities that go beyond these two functions, necessitating the convergence of ISAC with other system modules to enhance autonomy, adaptability, and sustainability.
\begin{table*}[!t]
\centering
\captionsetup{labelsep=newline, justification=centering}
\caption{Comparison of Different Level ISAC Designs with Various Requirements in LAWNs}
\label{table:ISAC_comparison}
\begin{tabular}{|c|c|c|c|c|}
\hline
\multirow{2}{*}{\diagbox[width=10em,height=2.26em]{\textbf{Requirements}}{\textbf{Category}}} 
  & \multicolumn{2}{c|}{\textbf{Node-Level ISAC}} 
  & \multicolumn{2}{c|}{\textbf{Network-Level ISAC}} \\
\cline{2-5}
  & Joint Waveform & SaC/CaS  
  & Centralized Fusion & Distributed Cooperation \\
\hline
Data Sharing 
  & None  
  & None  
  & Raw echoes  
  & Compressed data \\
\hline
Synchronization 
  & Not needed  
  & Not needed  
  & Tight  
  & Moderate \\
\hline
Performance Gain 
  & Integration gain  
  & Cooperation gain  
  & Fusion gain  
  & Collaborative gain \\
\hline
Latency 
  & / 
  & / 
  & Moderate  
  & Low-to-moderate \\
\hline
Scalability 
  & / 
  & /
  & Limited  
  & High \\
\hline
Complexity 
  & Low  
  & Low  
  & High  
  & Medium \\
\hline
\end{tabular}
\vspace{-0.5cm}
\end{table*}

\section{Beyond ISAC: Multi-Functional LAWN}
\begin{figure*} \vspace{-0.5cm}
    \centering
    \includegraphics[width=\linewidth]{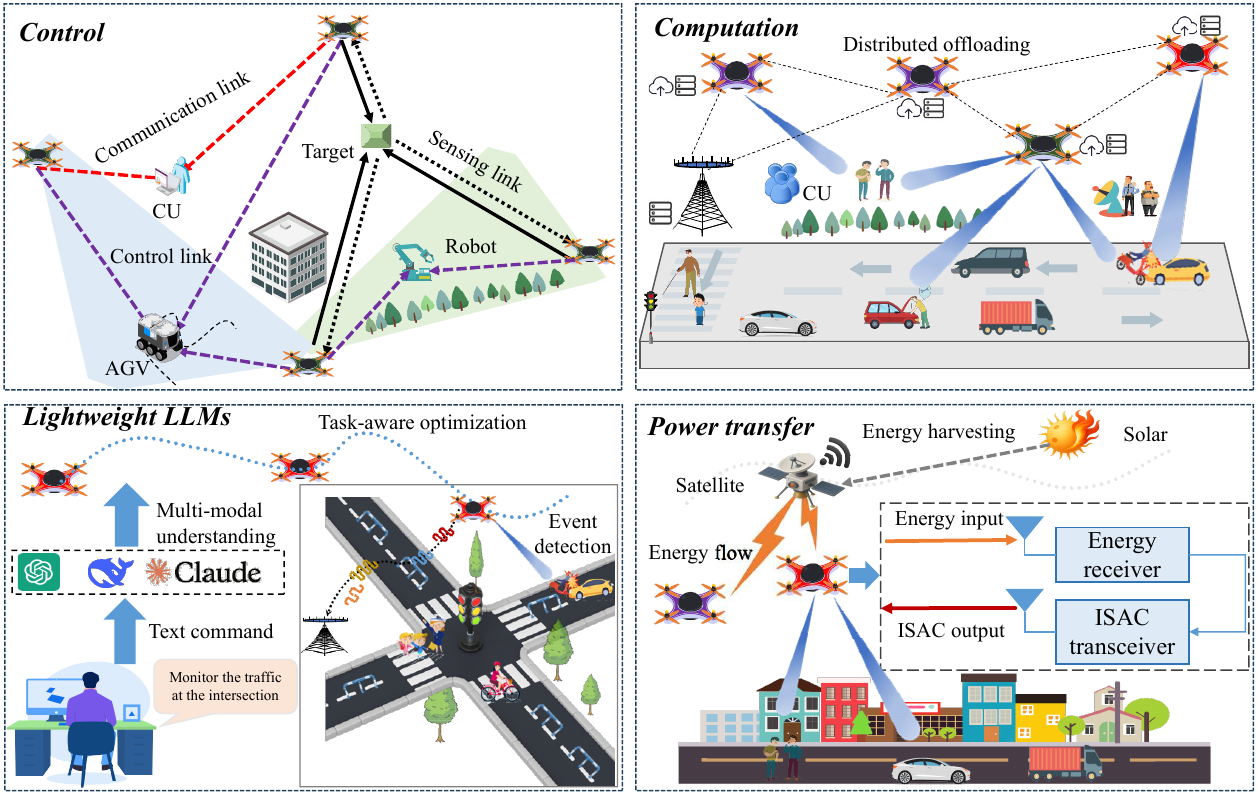}
    \caption{The interplay between ISAC and extended functions within LAWNs. \textbf{Wireless control }links ensure real-time coordination between air and ground. \textbf{Distributed offloading} enables efficient computation for large-scale LAWNs. \textbf{Lightweight LLMs} enable multi-modal understanding. \textbf{Wireless power transfer} ensures extended network lifetime. }
    \label{fig2:enter-label} 
    \vspace{-0.5cm}
\end{figure*}

In this section, we explore the evolution from ISAC-LAWN to multi-functional architectures, highlighting four key integration pathways depicted in Fig. \ref{fig2:enter-label}.
\subsection{Integrated Sensing, Communication, and Control}
The ultimate objective of intelligent systems deployed in LAWN is to execute precise control actions through timely and accurate interaction with dynamic environments. In such networks, control is not merely a functional extension of ISAC but a necessary step toward achieving closed-loop autonomy \cite{4118454,jin2025advancing}. As LAWN nodes operate under highly dynamic conditions, embedding control into the ISAC framework enables timely coordination, trajectory regulation, and safety assurance for aerial-ground agents. Within wireless control scenarios, the coupling between ISAC and control primarily arises through communication links that transmit sensing outcomes and control commands, which typically manifest in two fundamental configurations:

\vspace{0.5em}
\textit{1) Observation-based Local Control:} In this configuration, remote LAWN infrastructure nodes, such as BS and aerial monitoring platforms, conduct environmental sensing. The acquired information comprising object motion states is transmitted to the target agent, which then processes the received data locally to generate appropriate control decisions. This architecture is capable of supporting distributed mechanisms while alleviating computational load on the infrastructure. However, the overall control performance is critically dependent on the accuracy and timeliness of the transmitted observations. In particular, constraints on wireless resources, e.g., limited uplink bandwidth and non-negligible propagation delays, would give rise to a fundamental trade-off between the capacity of sensory data and the achievable control precision. Reducing the data volume may introduce quantization errors, whereas increasing transmission load may exacerbate congestion or latency. Revealing this trade-off is a paramount challenge in LAWN systems and necessitates further joint design concerning sensing, communication, and control.

\vspace{0.5em}
\textit{2) Communication-based Remote Control:} In this architecture, the remote infrastructure components within the LAWN  perform the control decision-making by relying on the
sensing results. Subsequently, the generated control commands are then transmitted to the controlled agents over downlink communications. Note that the effectiveness of this configuration is fundamentally constrained by the reliability of the downlink communication channel. Unlike observation-based schemes where sensing inaccuracies may be partially compensated through local feedback, command-based control entirely depends on the precise and timely reception of external control instructions.  Hence, reliable communications in terms of outage probability and bit error rate (BER) serve as critical metrics in quantifying the impact of wireless impairments on control fidelity. Moreover, the interplay between reliable communications and control-loop stability gives rise to a set of nontrivial trade-offs as well. For instance, reducing outage probability/BER often demands increased transmission power or redundancy, which in turn may affect system scalability and energy efficiency. As a consequence, a communication-control co-design framework is required, wherein physical-layer strategies may need to be explicitly optimized to meet control-level performance requirements.

\subsection{Integrated Sensing, Communication, and Computation}

Integrated sensing, communication, and computation (ISCC) form the core triad that supports intelligent and responsive interactions between aerial and terrestrial entities in LAWNs. By merging signal acquisition, data transmission, and task processing, ISCC systems address the stringent demands of latency, energy efficiency, and adaptability in dynamic environments. In these systems, LAWN agents, equipped with ISCC modules, not only capture real-time environmental data through radar or visual sensing but also transmit this information via shared spectrum links. Task execution is coordinated across hierarchical computing layers, ensuring that both local and offloaded tasks are processed efficiently. However, conventional ISCC frameworks encounter substantial limitations when applied to LAWN environments. Specifically, conventional ISCC systems mainly rely on centralized task offloading strategies, where drones usually offload computational tasks to fixed powerful BS or cloud servers, which may incur severe latency in case of large-scale resource-constrained LAWNs.  

To overcome these limitations, the integration of distributed peer-to-peer computation offloading further optimizes the ISCC framework within the LAWN. In this enhanced system, drones employ proximity-constrained offloading mechanisms, where tasks are offloaded exclusively to the nearest eligible drones, dynamically updating the adjacency matrix based on real-time distance measurements. Temporal resources are allocated through partitioned time slots that are subdivided into sequential offloading and computation phases, ensuring that computation follows data transfer. The system is designed to mitigate interference in D2D communication by employing orthogonal sub-channels, where each sub-channel is allocated to a single active link per drone. Queue-aware energy-latency balancing is incorporated, optimizing task arrivals, local processing, and drone offloading through dynamic queue backlogs.
To enhance coordination under decentralized conditions, distributed optimization techniques are employed, enabling individual agents to collaboratively solve global objectives such as energy minimization or latency balancing using only local information and limited neighbor communication. In light of primal-dual decomposition or alternating direction method of multipliers (ADMM)-based iterative methods, this approach ensures convergence to near-optimal configurations without relying on centralized control, thereby enhancing efficiency and robustness across D2D transmission, computation, and propulsion \cite{10812728}.

 \subsection{Integrated Sensing, Communication, and  Lightweight LLMs}

The convergence of lightweight LLMs with ISAC systems opens a transformative pathway toward semantic-level intelligence in LAWNs. While traditional onboard computation modules in LAWNs rely on rule-based or task-specific architectures to process sensor data, fuse information, and execute control commands, they often lack the flexibility and contextual awareness needed for mission-level autonomy. In contrast, LLMs possess emergent capabilities in semantic reasoning, temporal abstraction, and multimodal correlation, allowing LAWN systems to move beyond signal-level processing into high-level understanding and decision-making.

In a typical workflow, ISAC-enabled LAWN agents acquire multi-modal observations via shared transceiver hardware, including radar profiles,  communication logs, and task metadata, which are processed by nearby edge servers equipped with resource-efficient LLMs. Leveraging extensive pretraining on diverse datasets, the LLM interprets the semantic intent behind the observed environment, identifies latent inter-agent dependencies, and formulates high-level guidance. Taking an example, as shown in Fig. \ref{fig2:enter-label}, a user may raise a request to capture images of the traffic accident site, such that goal-aware trajectory planning, coordinated spectrum access, and anomaly response need to be performed by lightweight LLMs. More importantly, the feedback from the LLM also influences physical-layer ISAC operations.  Semantic tokens generated by the LLMs can be used to dynamically adjust waveform parameters to prioritize S\&C missions  \cite{10643253}.
Such integration also offers scalability and generalization. LLMs can adapt to new environments and mission requirements with minimal retraining, providing a unified backbone across surveillance, delivery, disaster relief, and collaborative airspace management applications. Moreover, the ability to abstract over noisy or incomplete sensory input makes LLMs particularly well-suited for real-world LAWN deployments, where unpredictable weather, topology, or interference may degrade raw signal quality. In essence, embedding LLMs into the ISAC-LAWN framework fundamentally redefines the system’s architecture toward a more interpretable, robust, and mission-aware network, paving the way for human-level cognitive collaboration and autonomous decision-making in the low-altitude ecosystem.

 \subsection{Integrated Sensing, Communication, and Power Transfer}

As LAWNs continue to evolve toward prolonged operation and infrastructure independence, energy sustainability becomes a key bottleneck. Plugging WPT into the ISAC framework offers a promising solution that allows LAWN nodes to sense the environment, exchange information, and replenish energy through shared radio resources \cite{8476597,6951347}. In this context, electromagnetic signals are no longer only transmitting communication data or sensing information, but also delivering wireless energy, thus enabling unified waveform designs that simultaneously support threefold functionality. Note that the joint ISAC-powering design necessitates trade-offs between spectral efficiency, sensing precision, and energy transfer efficiency. For example, narrow-beam, high-power transmissions may improve wireless charging range but degrade sensing coverage due to reduced waveform diversity. Similarly, energy-oriented beamforming may lead to alignment mismatches with optimal sensing directions or communication paths, particularly in multi-agent LAWN environments where mobility and topology dynamics introduce spatiotemporal variability. More interestingly, satellites and high-altitude platforms (HAPs) can be further leveraged to harvest solar energy and wirelessly charge LAWNs, thereby extending network lifetime and enhancing safety compared to conventional ground-based WPT mechanisms. Additionally, practical challenges such as non-linear energy harvesting characteristics, rectifier hardware constraints further complicate the integration process. 
  \begin{figure*}[h]
      \centering 
      \includegraphics[width=0.95\linewidth]{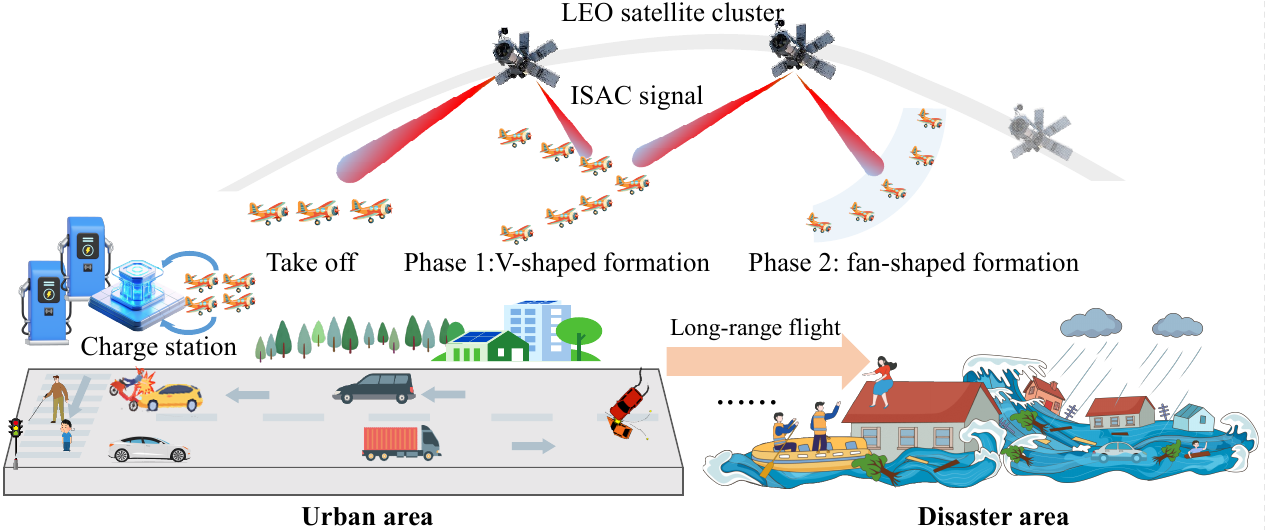}
      \caption{Case study of task-aware formation control within a LAWN-enabled emergency response scenario.}
      \vspace{-0.5cm} \label{case-sce}
      \end{figure*}
\subsection{Lessons Learned}
The evolution toward multi-functional LAWNs underscores that integrating control, computation, intelligence, and power into ISAC frameworks creates both profound synergies and inherent tensions. Success hinges on transcending traditional domain-specific design in favor of a unified, cross-functional architecture that dynamically balances competing objectives to meet the holistic demands of next-generation aerial missions.

\begin{figure*}[h]
    \centering
    \begin{subfigure}{0.32\linewidth}
        \includegraphics[width=\linewidth]{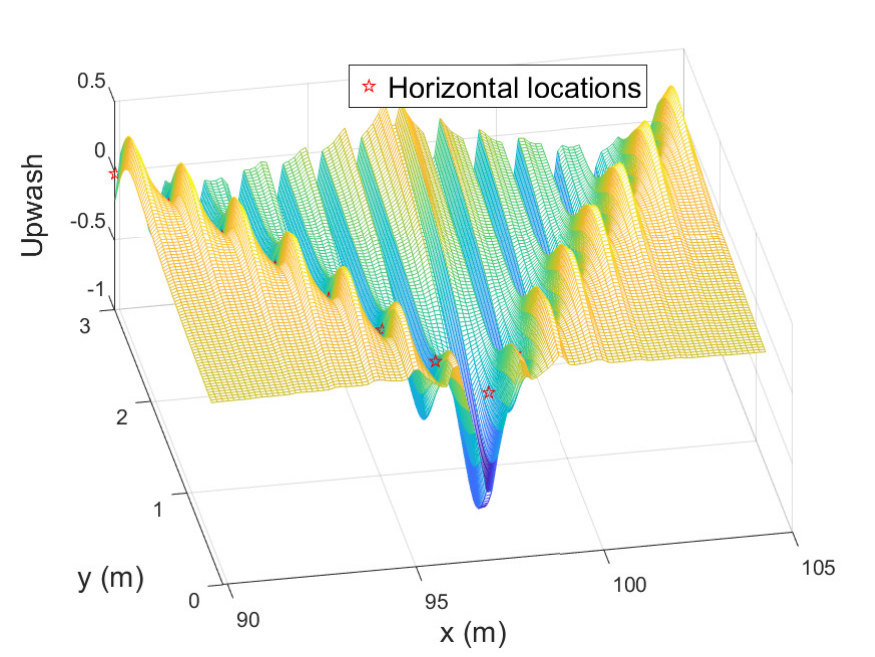}
        \caption{V-shaped formation and aerodynamics.}
        \label{fig:vshape}
    \end{subfigure}
    \hfill
    \begin{subfigure}{0.32\linewidth}
        \includegraphics[width=\linewidth]{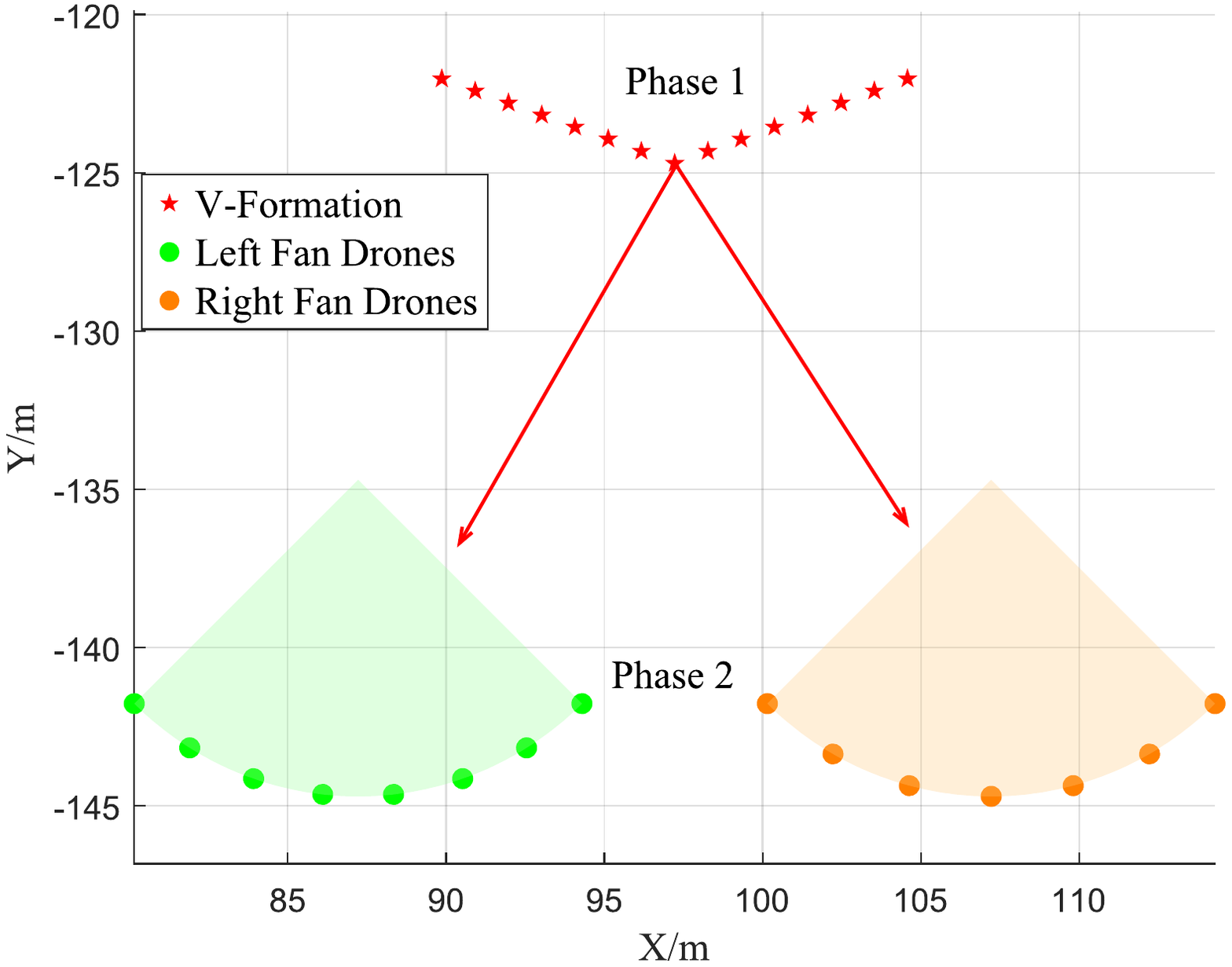}
        \caption{Fan-shaped drones formation.}
        \label{fig:fan}
    \end{subfigure}
    \hfill
    \begin{subfigure}{0.32\linewidth}
        \includegraphics[width=\linewidth]{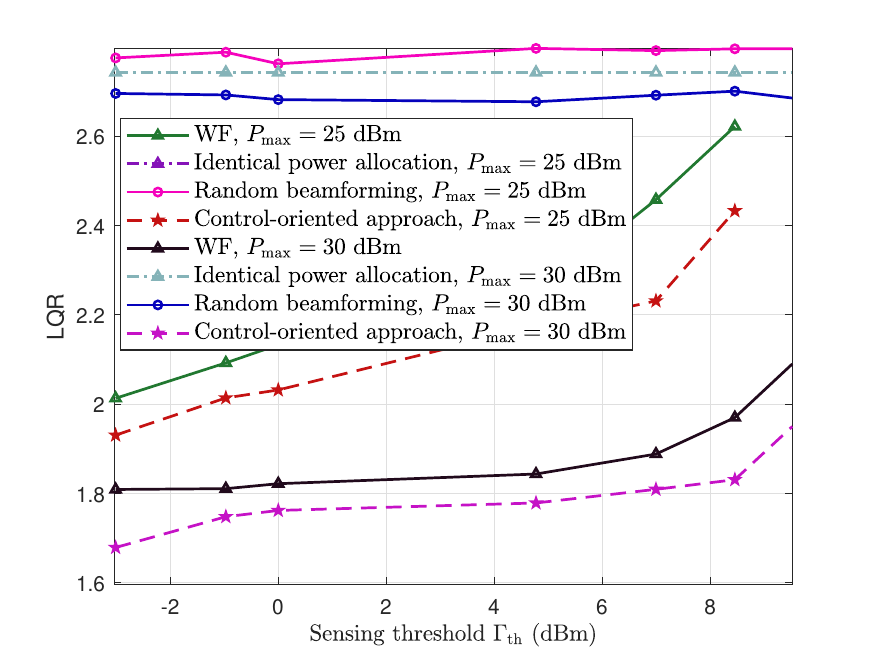}
        \caption{LQR cost under different sensing conditions.}
        \label{fig:lqr}
    \end{subfigure}
    \caption{Two different formation shapes in two distinct phases and the control performance of LQR cost with different sensing requirements.}
    \label{fig:formation_overall}
\end{figure*}
\section{Case Study: Task-Aware Formation Control }
\subsection{System Description}
Based on the above discussion, we consider a representative LAWN-enabled emergency response scenario. In this section,  a fleet of fixed-wing drones is dispatched from a centralized charging hub in the urban area to perform real-time survivor search across a remote mountainous region lacking terrestrial infrastructure. One drone is specially designated as the leader of this fleet. A low Earth orbit (LEO) satellite cluster acts as an ISAC base station, performing environmental observations of the target area and transmitting control commands to the fleet leader throughout the mission. Moreover, the satellite can harvest energy from the solar, followed by performing WPT to the fleet to extend survival time. The system scenario is shown in Fig. \ref{case-sce}. The entire mission consists of two distinct phases, each associated with evolving control objectives and task-aware formation shaping: 

\textbf{1) Energy-Saving Formation Flight:}
In the first phase, the swarm initiates a long-range flight toward a disaster site. During this phase, energy saving is prioritized. To this end, the fleet adopts a self-organized flight pattern guided by aerodynamic interaction principles.  Specifically,  when fixed-wing UAVs fly in proximity, they will generate two vortices around the wingtip, capable of providing upward induced airflow (upwash) that can alleviate the aerodynamic drag for a follower drone entering this zone. As a consequence, by positioning trailing UAVs within the upwash region behind the preceding UAV, additional lift can be passively acquired, thereby allowing lower thrust for level flight. 

\textbf{2) Maximum Exploring Coverage Formation:}
Upon approaching the mission area, the focus transitions from energy conservation to situational awareness and spatial coverage. Guided by the satellite’s reconnaissance feedback and terrain constraints, the formation reconfigures from its prior aerodynamic alignment into a fan-shaped arrangement, expanding laterally toward the region of interest. This structure allows the drones to enable simultaneous multi-angle observation and minimize blind zones, improving search efficiency and detection probability while preserving coordination and flight stability.

 To guarantee the formation success within the whole mission period, the LEO satellite senses the terrain profile and tracks the fleet’s current kinematic states continuously. Based on the observed trajectories and environmental conditions, it generates appropriate control commands and transmits them to the drone formation. The ultimate goal is to improve wireless control performance in terms of linear quadratic regulator (LQR) cost.
\subsection{Result Discussions}
We consider a mission scenario within a $1\,\text{km} \times 1\,\text{km}$ square area, where $15$ fixed-wing drones operate at a constant altitude of $50$\,m. Each drone is characterized by a wingspan of $1$\,m and a cruising speed of $5$\,m/s. The proposed task-aware formation strategy is evaluated under a two-phase operational paradigm.
In \textbf{Phase 1}, the drones exhibit a V-shaped spatial arrangement driven by aerodynamic interactions, as illustrated in Fig.~\ref{fig:vshape}. Specifically, under the influence of the upwash effect generated by the leading drones, trailing drones are naturally attracted to the energetically favorable regions, leading to a self-organized V-formation and thereby contributing to improved collective endurance during long-range flight, which is exactly aligned with the phenomenon of bird flock immigration.
As the swarm approaches the task-relevant region, it transitions into \textbf{Phase 2}, adopting a symmetric fan-shaped formation centered along the negative $y$-axis, as shown in Fig.~\ref{fig:fan}. This reconfiguration enhances spatial coverage and sensing diversity, which are essential for precise perception and situational response near the disaster zone. Fig.~\ref{fig:lqr} presents the wireless control performance in terms of LQR cost under various sensing thresholds $\Gamma_\mathrm{th}$ across different ISAC-beamforming design strategies at the LEO. One can observe that the proposed control-oriented approach consistently achieves lower LQR cost compared with baseline schemes. It is evident that the simulations validate the efficacy of ISAC-supported formation design, where the drones demonstrate enhanced energy efficiency, spatial awareness, and control robustness in dynamic mission scenarios.

\section{Open Challenges and Future Directions}

\subsection{Low-Altitude Uncrewed Aircraft Traffic Management}

Managing dense low-altitude airspace requires global situational awareness and real-time regulation, which can be achieved through network-level ISAC. By enabling multiple infrastructure nodes and aerial agents to collaboratively fuse their sensing and communication data, network-level ISAC supports distributed environmental map construction and macroscopic fundamental diagram (MFD)-based airspace modeling. Through continuous estimation of drone density, flow rate, and congestion levels, network-level ISAC facilitates adaptive zoning and dynamic signalization to enhance global efficiency and operational safety. By jointly coordinating sensing coverage and data dissemination across the network, ISAC supports enforcement of capacity constraints and stabilization of flow dynamics, laying the groundwork for scalable and autonomous traffic management.

\subsection{Trustworthy Identity Authentication}

In congested LAWN environments, reliable identification of aerial and ground agents is critical for secure and coordinated operations. This challenge can be addressed via node-level ISAC, where each UAV or ground node exploits its integrated sensing and communication functions to map physical-layer signatures such as Doppler shifts, time-frequency features, and RF fingerprints. By further incorporating sensing-derived behavioral cues, node-level ISAC enables robust identity mapping, reducing dependency on digital credentials and enhancing resistance to spoofing and impersonation attacks. 
 \subsection{3D Interference Pattern Prediction}
At the node level, ISAC transceivers struggle to cope with interference from communication, sensing, control, and WPT links that occupy different ranges, i.e., long range for communication, medium for sensing, and short for power transfer. At the network level, overlapping beams and mobility make the 3D spatial distribution of interference highly dynamic and difficult to coordinate. A key research direction is to construct 3D interference radio maps by combining model-based long-term predictions with data-driven short-term updates.




\section{Conclusion}

This article has explored the evolution of LAWNs toward a multi-functional architecture that extends beyond traditional ISAC, synergizing control, computation, WPT, and LLM-based decision-making. We have examined the distinct roles of node-level and network-level ISAC designs within LAWN systems.  We have also presented a representative case study to demonstrate how these functions collectively operate in LAWN-enabled applications, where simulations are conducted to validate the effectiveness of joint design. Finally, we have highlighted several future research directions for scalable and adaptive LAWN systems in dynamic environments.

\bibliographystyle{ieeetr}
\bibliography{reference}

\end{document}